\UseRawInputEncoding

\documentclass[twocolumn]{aastex63}

\usepackage{amsmath}
\usepackage{epsfig}
\usepackage{amssymb}
\usepackage{graphicx}
\usepackage{color}
\usepackage{natbib}

\def\gtaprx {\lower .1ex\hbox{\rlap{\raise .6ex\hbox{\hskip .3ex
	{\ifmmode{\scriptscriptstyle >}\else
		{$\scriptscriptstyle >$}\fi}}}
	\kern -.4ex{\ifmmode{\scriptscriptstyle \sim}\else
		{$\scriptscriptstyle\sim$}\fi}}}
\def\ltaprx {\lower .1ex\hbox{\rlap{\raise .6ex\hbox{\hskip .3ex
	{\ifmmode{\scriptscriptstyle <}\else
		{$\scriptscriptstyle <$}\fi}}}
	\kern -.4ex{\ifmmode{\scriptscriptstyle \sim}\else
		{$\scriptscriptstyle\sim$}\fi}}}

\newcommand{\cutt}[1]{\textcolor{blue}{}}

\newcommand{\Ms}{{\ensuremath{M_{\odot} }}}

\newcommand{\C}{{\ensuremath{^{12}\mathrm{C}}}}

\newcommand{\N}{{\ensuremath{^{14} \mathrm{N}}} }

\accepted{by ApJL on November 1, 2021}

\shorttitle{DCBH Radio Flux}
\shortauthors{Whalen et al.}

\begin{document}

\title{Radio Power from Direct-Collapse Black Holes}

\correspondingauthor{Daniel J. Whalen}
\email{daniel.whalen@port.ac.uk}

\author{Daniel J. Whalen}
\affiliation{Institute of Cosmology and Gravitation, University of Portsmouth, Portsmouth PO1 3FX, UK}
\affiliation{Ida Pfeiffer Professor, University of Vienna, Department of Astrophysics, Tuerkenschanzstrasse 17, 1180, Vienna, Austria}

\author{Mar Mezcua}
\affiliation{Institute of Space Sciences (ICE, CSIC), Campus UAB, Carrer de Magrans, 08193 Barcelona, Spain}
\affiliation{Institut dÕEstudis Espacials de Catalunya (IEEC), Carrer Gran Capit\'a, 08034 Barcelona, Spain }

\author{Samuel J. Patrick}
\affiliation{Institute for Astronomy, University of Edinburgh, Blackford Hill, Edinburgh\ EH9\ 3HJ, UK}

\author{Avery Meiksin}
\affiliation{Institute for Astronomy, University of Edinburgh, Blackford Hill, Edinburgh\ EH9\ 3HJ, UK}

\author{Muhammad A. Latif}
\affiliation{Physics Department, College of Science, United Arab Emirates University, PO Box 15551, Al-Ain, UAE}




\begin{abstract}

Direct-collapse black holes (DCBHs) forming at $z \sim$ 20 are currently the leading candidates for the seeds of the first quasars, over 200 of which have now been found at $z >$ 6.  Recent studies suggest that DCBHs could be detected in the near infrared by the {\em James Webb Space Telescope}, {\em Euclid}, and the {\em Roman Space Telescope}.  However, new radio telescopes with unprecedented sensitivities such as the Square Kilometer Array (SKA) and the Next-Generation Very Large Array (ngVLA) may open another window on the properties of DCBHs in the coming decade.  Here we estimate the radio flux from DCBHs at birth at $z =$ 8 - 20 with several fundamental planes of black hole accretion.  We find that they could be detected at $z \sim$ 8 by the SKA-FIN all-sky survey.  Furthermore, SKA and ngVLA could discover 10$^6$ - 10$^7$ \Ms\ BHs out to $z \sim$ 20, probing the formation pathways of the first quasars in the Universe.
 
\end{abstract}


\section{Introduction}

Direct-collapse black holes (DCBHs) are one of the leading contenders for the origin of the first quasars in the Universe, over 200 of which have now been found at $z >$ 6 \citep{mort11,ban18,mats19,wang21}.  They are thought to form when rapidly-accreting supermassive primordial stars in atomically-cooling halos collapse at masses of a few 10$^4$ - 10$^5$ \Ms\ at $z \sim$ 20 \citep{tyr17,tyr21a,herr21a}.  DCBHs may be the seeds of the first supermassive black holes (SMBHs) because it is difficult for ordinary Population III (Pop III) star BHs to grow rapidly after birth \citep{wan04,wf12,srd18}.  DCBHs are born with much larger masses and in much higher densities in host halos that can retain their fuel supply, even when it is heated by X-rays (\citealt{jet13} -- see \citealt{titans} for recent reviews). 

A number of studies have examined the prospects for detection of supermassive Pop III stars \citep{sur18a,sur19a} and DCBHs \citep{pac15,bar18,wet20b} in the near infrared (NIR) by the {\em James Webb Space Telescope} ({\em JWST}), {\em Euclid}, and the {\em Roman Space Telescope} ({\em RST}).  They found that DCBHs could be detected by {\em JWST} at $z \lesssim$ 20 and by {\em Euclid} and {\em RST} at $z \lesssim$ 6 - 8, although lensing by galaxy clusters and massive galaxies in their wide fields could extend these detections up to $z \lesssim$ 10 - 15 \citep{vik21a}.  They can be distinguished from primordial galaxies in color-color space in appropriate NIR filters and from cool, dim foreground stars because they are transients due to fluctuations in accretion flows.  DCBHs may also be detected out to $z \sim$ 10 by future X-ray missions such as the {\em Advanced Telescope for High-Energy Astrophysics} ({\em ATHENA}) and {\em Lynx}.

\begin{figure*} 
\begin{center}
\begin{tabular}{cc}
\epsfig{file=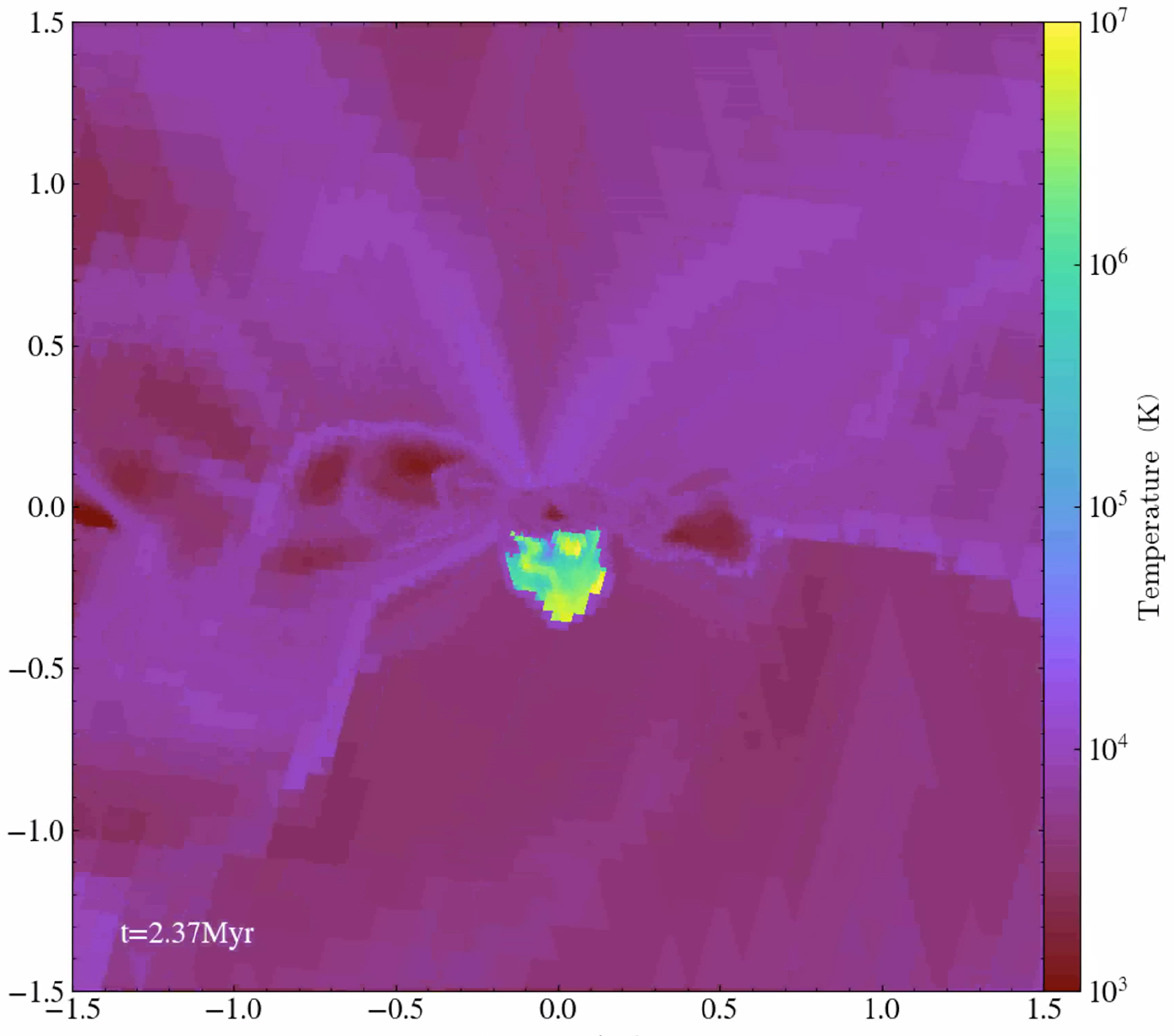,width=0.55\linewidth,clip=}  &
\epsfig{file=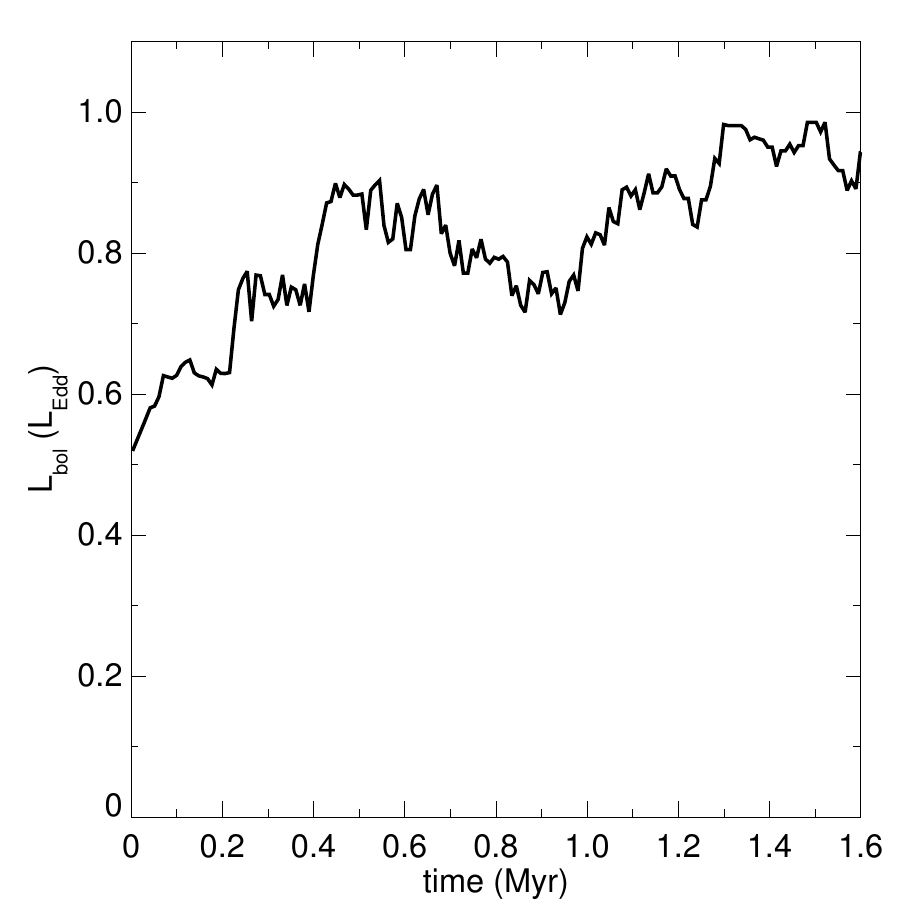,width=0.4\linewidth,clip=}  \\
\end{tabular}
\end{center}
\caption{Left:  temperature image of an atomically-cooled disk surrounding a DCBH 2.37 Myr after birth.  An ultracompact \ion{H}{2} region (yellow/green) with a radius of $\sim$ 0.15 pc is visible below the midplane of the disk.  The X-rays are trapped in the disk because of large ambient densities and the ram pressure of infalling gas.  The image is 3 pc on a side.  Right:  DCBH bolometric luminosity in units of the Eddington rate.}
\vspace{0.1in}
\label{fig:disk} 
\end{figure*}

\begin{figure*} 
\begin{center}
\begin{tabular}{cc}
\epsfig{file=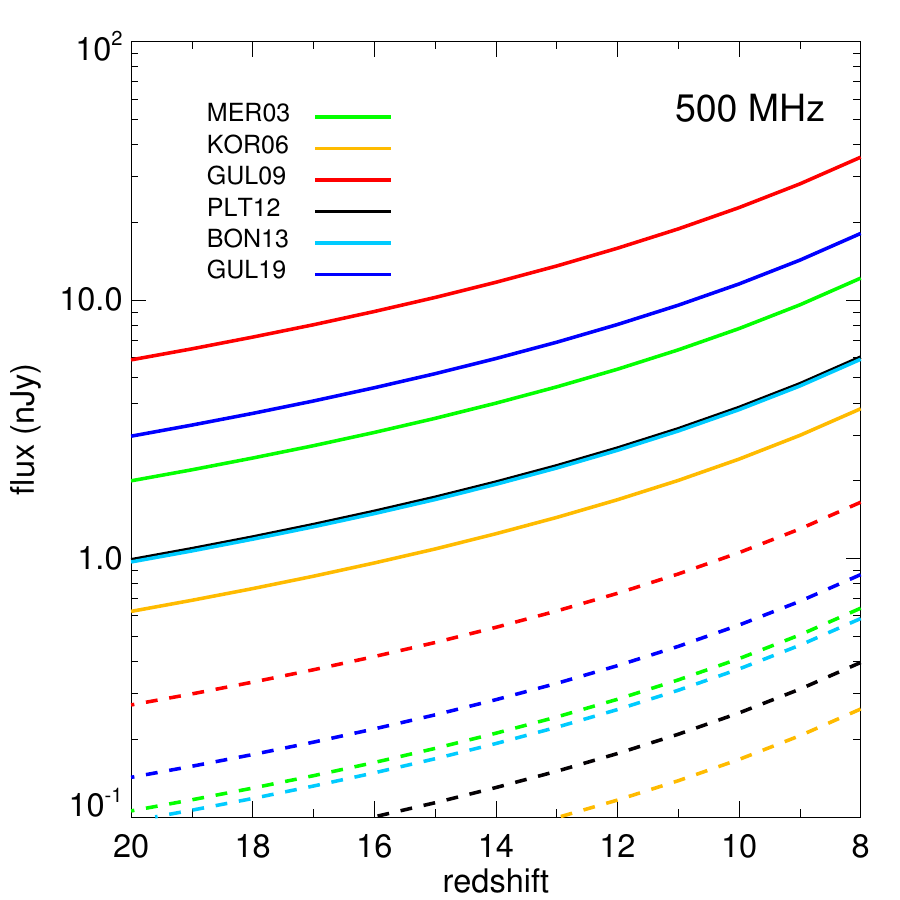,width=0.35\linewidth,clip=}  &
\epsfig{file=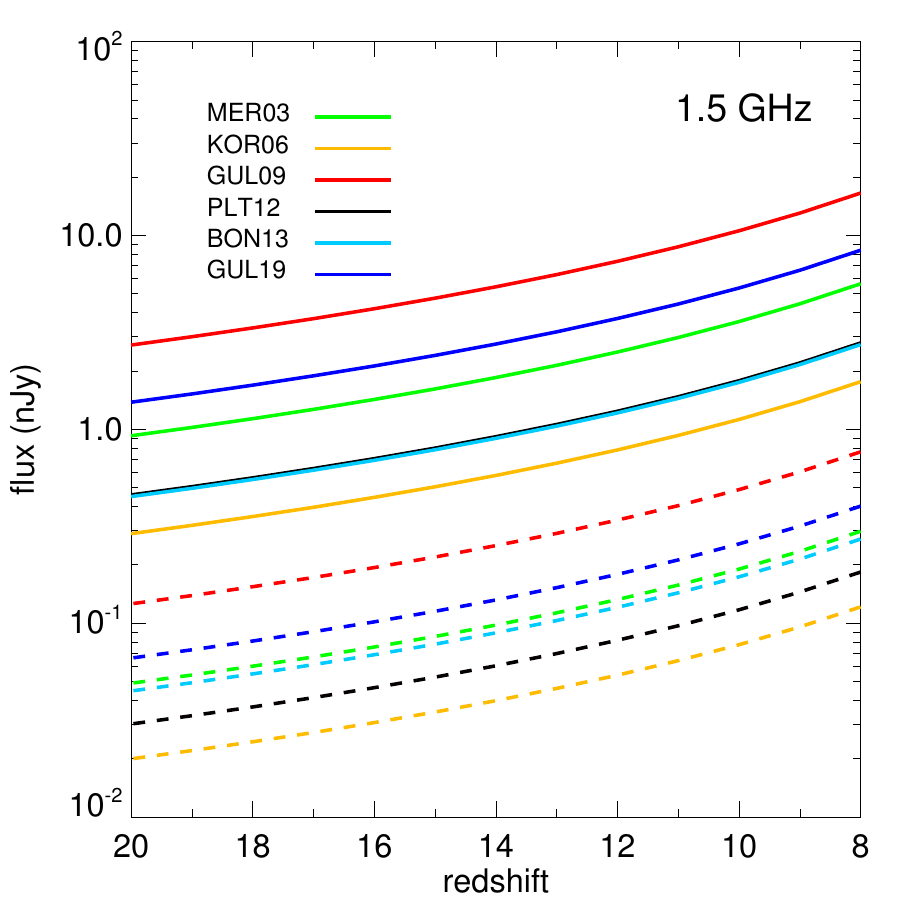,width=0.35\linewidth,clip=}  \\
\epsfig{file=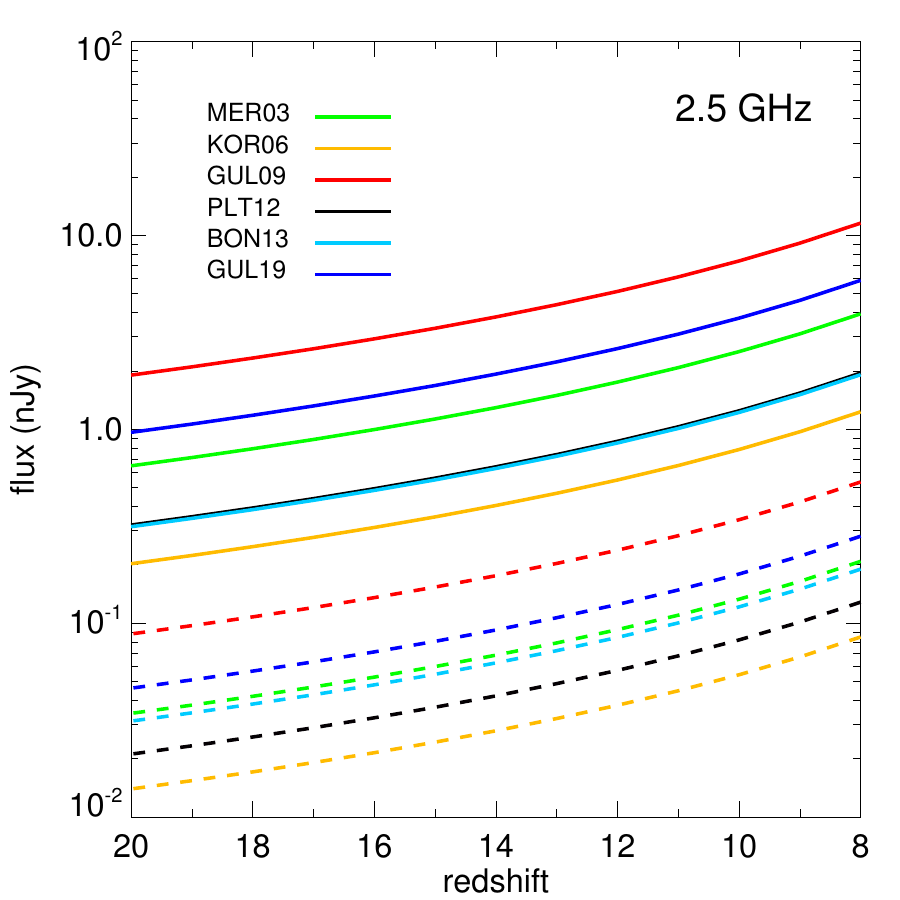,width=0.35\linewidth,clip=}  &
\epsfig{file=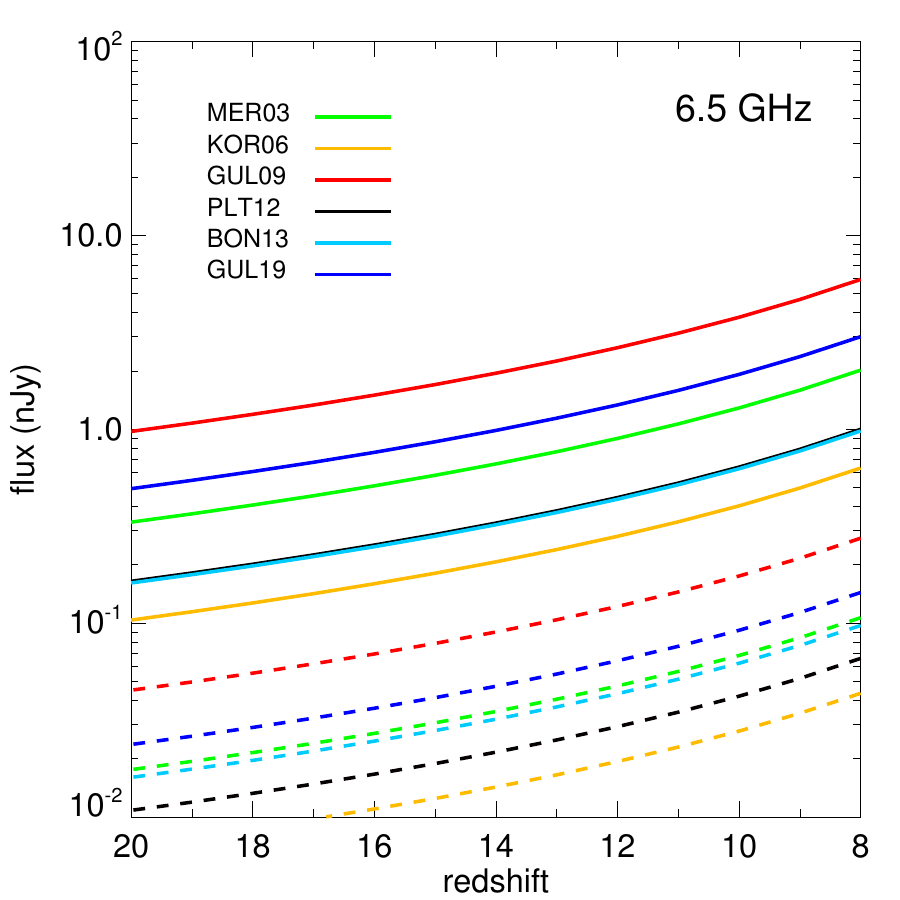,width=0.35\linewidth,clip=}  
\end{tabular}
\end{center}
\caption{Radio fluxes for 10$^5$ \Ms\ (dashed) and 10$^6$ \Ms\ (solid) DCBHs at $z =$ 8 - 20 from 6 FPs.  Upper left:  500 MHz.  Upper right:  1.5 GHz.  Lower left:  2.5 GHz.  Lower right:  6.5 GHz.}
\vspace{0.1in}
\label{fig:flux} 
\end{figure*}

DCBHs are also expected to be radio sources, and they could be detected by a new generation of observatories with unprecedented sensitivities such as the Square Kilometer Array (SKA) and the next-generation Very Large Array (ngVLA) in the coming decade.  Recent estimates of radio power from a DCBH in the Ly$\alpha$ emitter CR7 at $z =$ 6.6 indicate that it could reach $\sim$ 100 nJy at 1.0 GHz, well above the eventual 20 nJy detection limit planned for the SKA-FIN all-sky survey \citep{wet20a}.  Nascent DCBHs would be free of competing sources of radio emission in their host halos that are present in CR7, such as young supernova remnants \citep[SNRs;][]{mw12} and \ion{H}{2} regions due to star formation \citep{con92}, because they form in halos before other stars.  

\citet{yf21} recently estimated DCBH radio fluxes under the assumption that they drive strong jets and found that they should be detectable by the SKA and ngVLA at $z \sim$ 10, depending on the inclination angle of the jet.  However, steady jets have only been observed in active galactic nuclei with luminosities $L \lesssim 0.01 \, L_{\mathrm{Edd}}$, where $L_{\mathrm{Edd}}$ is the Eddington limit, and intermittent jets have been seen in quasars at $L \sim L_{\mathrm{Edd}}$ \citep{mh08}.  At these accretion rates the disk is geometrically thick and radiatively inefficient.  No jets have been found at 0.01 $L_{\mathrm{Edd}} < L < L_{\mathrm{Edd}}$, when the disk is geometrically thin and radiatively efficient.  Recent radiation hydrodynamical simulations of the formation of a $z =$ 7.1 quasar found that the DCBH accretes at 20\% - 80\% of the Eddington limit at birth, a regime in which no jets are expected to form, but they could not resolve the accretion disk of the BH \citep{smidt18}.  

We have instead applied fundamental planes of BH accretion to estimate DCBH radio fluxes for a range of redshifts.  The accretion rates in these planes are taken from a new cosmological radiation hydrodynamical simulation that resolves flows onto the DCBH on scales of 0.01 pc that have never before been achieved.  In Section 2 we discuss our DCBH accretion simulation and radio flux calculations.  We calculate radio emission as a function of DCBH mass and redshift and estimate the ngVLA integration times required to detect them in Section 3.  We also determine the minimum DCBH masses required for detection by currently planned SKA and ngVLA surveys as a function of redshift and then conclude in Section 4. 

\section{Numerical Method}

We estimate radio fluxes from 10$^5$ \Ms\ and 10$^6$ \Ms\ DCBHs at $z =$ 8 - 20.  We adopt 10$^6$ \Ms\ as a firm upper limit to the mass of a DCBH because stellar evolution calculations predict that such objects could form in the accretion rates found in atomically-cooled halos \citep[e.g.,][]{um16} and because some of these halos collapse at $\sim$ 10$^8$ \Ms\ instead of 10$^7$ \Ms\ \citep[see Table 1 of][]{pat21a}.  Alternatively, it could represent a partially evolved DCBH that has grown in mass.  We extract DCBH accretion rates from a cosmological radiation hydrodynamics simulation and use them to estimate DCBH radio fluxes in six fundamental planes of BH accretion.  We also estimate the flux from the ultracompact \ion{H}{2} region formed by X-rays from the BH to determine if it is an important source of radio emission.

\subsection{DCBH Bolometric Luminosities}

Our DCBH simulation is performed with the Enzo adaptive mesh refinement (AMR) cosmology code \citep{enzo}.  Our simulation box is 1.5 $h^{-1}$ Mpc with a 256$^3$ root grid and three nested grids centered on the halo for an effective initial resolution of 2048$^3$.  We initialize the run at $z =$ 200 with Gaussian random perturbations calculated with MUSIC \citep{hahn11} and allow up to 15 levels of AMR after the onset of atomic cooling in the halo for a maximum resolution of 0.014 pc.  We use second-year \textit{Planck} cosmological parameters:  $\Omega_{\mathrm M} = 0.308$, $\Omega_\Lambda = 0.691$, $\Omega_{\mathrm b}h^2 = 0.0223$, $\sigma_8 =$ 0.816, $h = $ 0.677 and $n =$ 0.968 \citep{planck2}.

We include six-species nonequilibrium primordial gas chemistry (H, He, e$^-$, H$^+$, He$^+$, H$^{2+}$) without H$_2$ to ensure isothermal cooling and collapse.  X-rays from the DCBH are propagated in the box with the MORAY radiation transport package \citep{moray}.  Ionizations and Compton heating due to X-rays and secondary ionizations by energetic photoelectrons are included in the chemistry and gas energy equations along with the usual primordial cooling processes: collisional excitational and ionizational cooling by H and He, recombinational cooling, bremsstrahlung cooling, and inverse Compton cooling by the cosmic microwave background (CMB).   We use an alpha disk model \citep{alphad} to tally accretion rates for the DCBH because it has a subgrid prescription for angular momentum transport out of the center of the disk, which our simulation does not fully resolve.  The bolometric luminosity in Enzo is
\begin{equation}
L_{\mathrm{bol}} = \epsilon \dot{m_{\mathrm{BH}}} c^2, 
\end{equation}
where the mean radiative efficiency $\epsilon =$ 0.1 and $\dot{m_{\mathrm{BH}}}$ is the accretion rate.  We take $L_{\mathrm{bol}}$ to be entirely 1 keV photons  \citep{smidt18}. 

The halo in our model begins to atomically cool when it reaches a mass of 1.44 $\times$ 10$^7$ \Ms\ at $z =$ 17 and forms a massive disk at its center.  At this point we turn on 1 keV X-rays from a 10$^5$ \Ms\ DCBH at the center of the disk and evolve it for 1.66 Myr.  We show an edge-on view of the disk in the left panel of Figure~\ref{fig:disk} at 2.37 Myr.  It is approximately 3 pc in diameter, and the nascent ultracompact (UC) \ion{H}{2} region of the BH is visible as the lobe of 5 $\times$ 10$^6$ K ionized gas extending $\sim$ 0.5 pc below the midplane of the disk at its center.  Radiation from the BH remains trapped deep in the disk throughout the run because of gas densities that exceed 10$^{10}$ cm$^{-3}$ and large ram pressures due to heavy infall that can reach 1 \Ms\ yr$^{-1}$.  We plot the bolometric luminosity, $L_{\mathrm{bol}}$, for the DCBH in the right panel of Figure~\ref{fig:disk}.  There are fluctuations due to turbulent flows in the disk but they average $\sim$ 0.85 $L_\mathrm{Edd}$.  We adopt this luminosity for a 10$^5$ \Ms\ DCBH in our fundamental plane calculations, where $L_\mathrm{Edd} =$ 1.26 $\times$ 10$^{38}$ ($M/$\Ms) erg s$^{-1}$. We also use this accretion rate for the 10$^6$ \Ms\ DCBH because it is consistent with those for partially evolved DCBHs at similar masses in other simulations \citet{smidt18,latif20b}.  

\begin{figure*} 
\begin{center}
\begin{tabular}{cc}
\epsfig{file=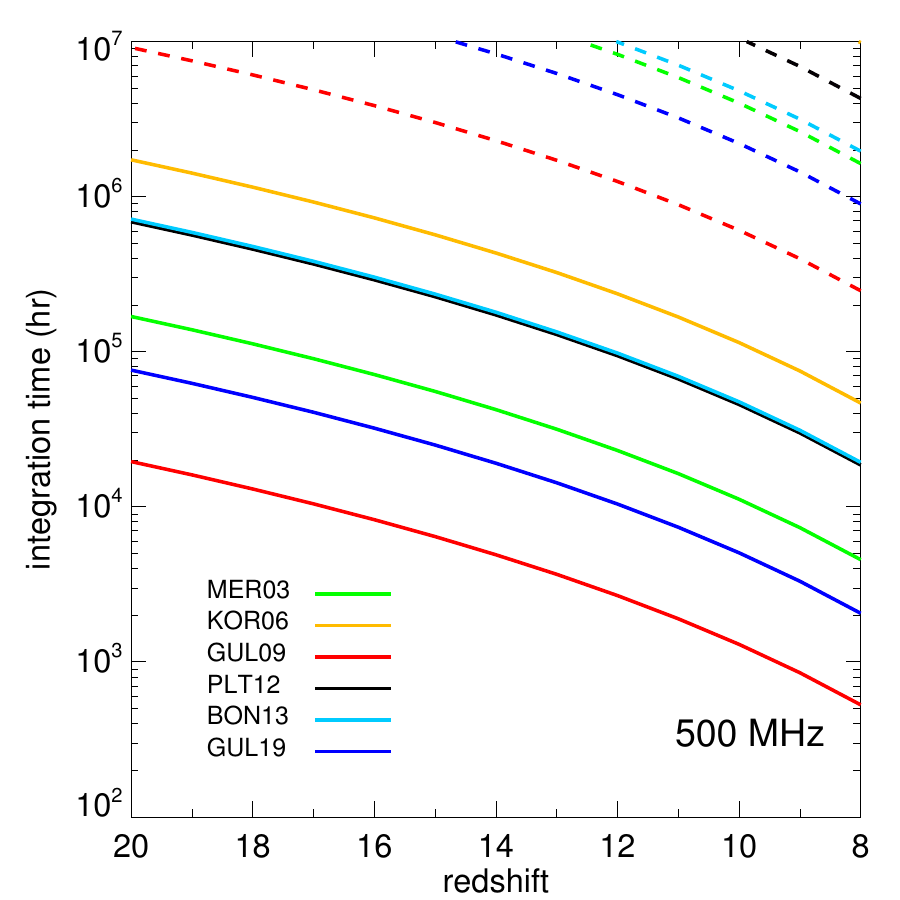,width=0.35\linewidth,clip=}  &
\epsfig{file=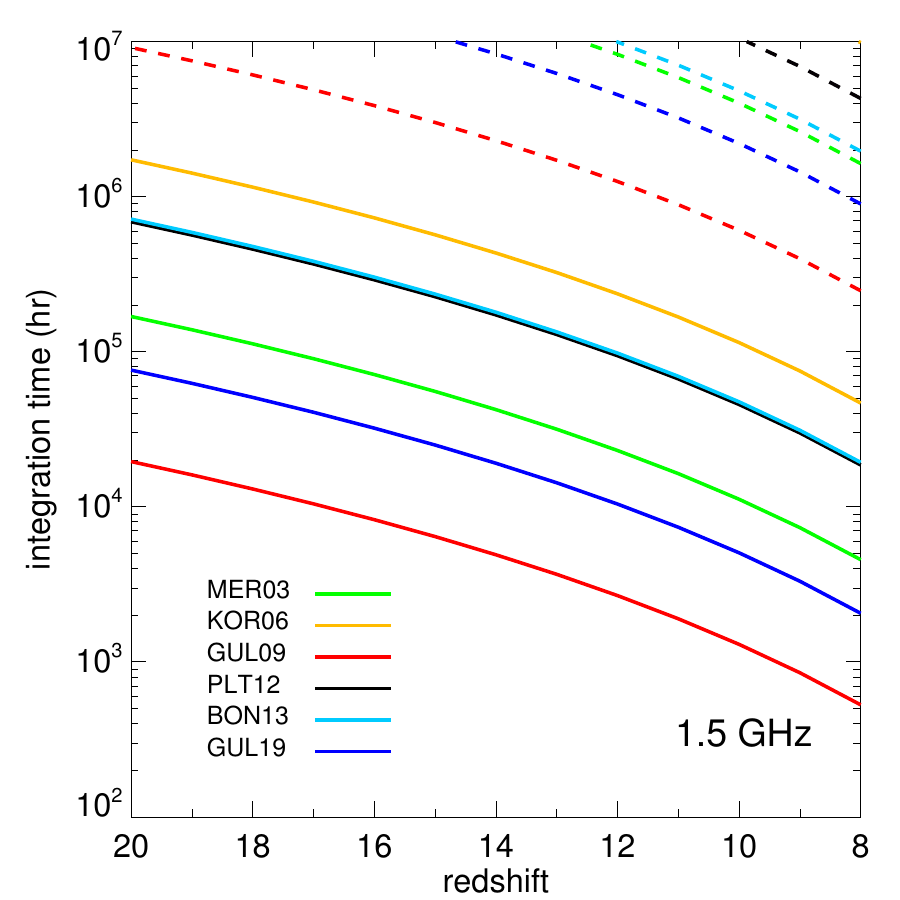,width=0.35\linewidth,clip=}  \\
\epsfig{file=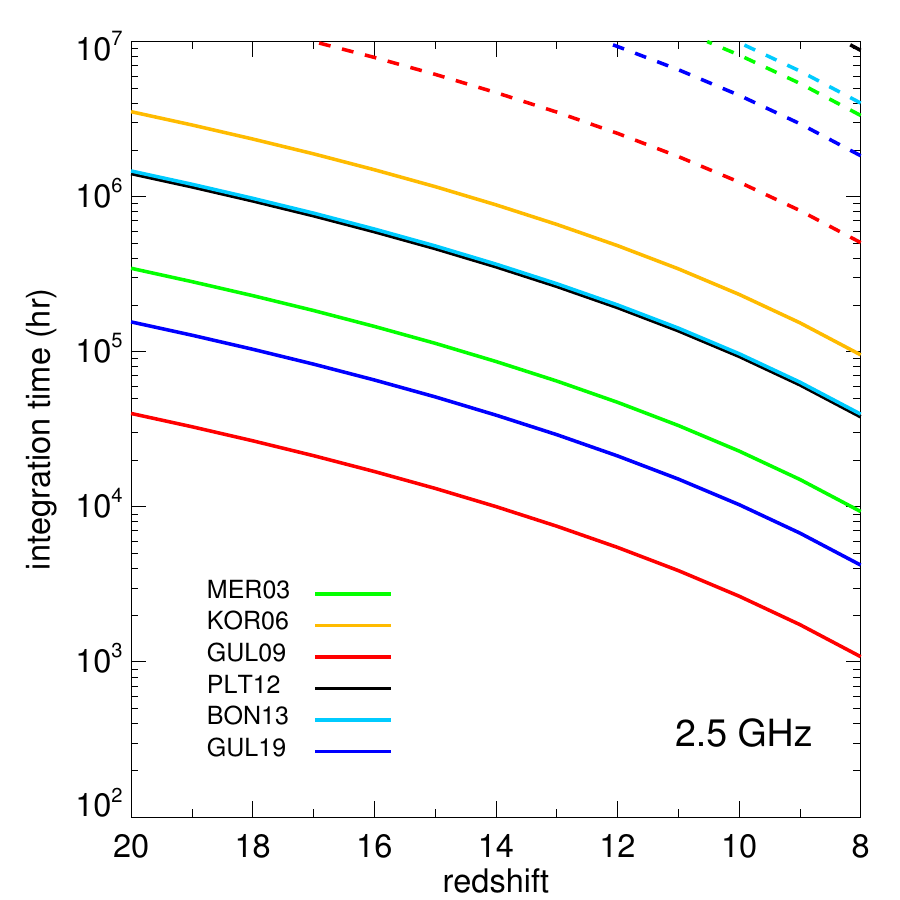,width=0.35\linewidth,clip=}  &
\epsfig{file=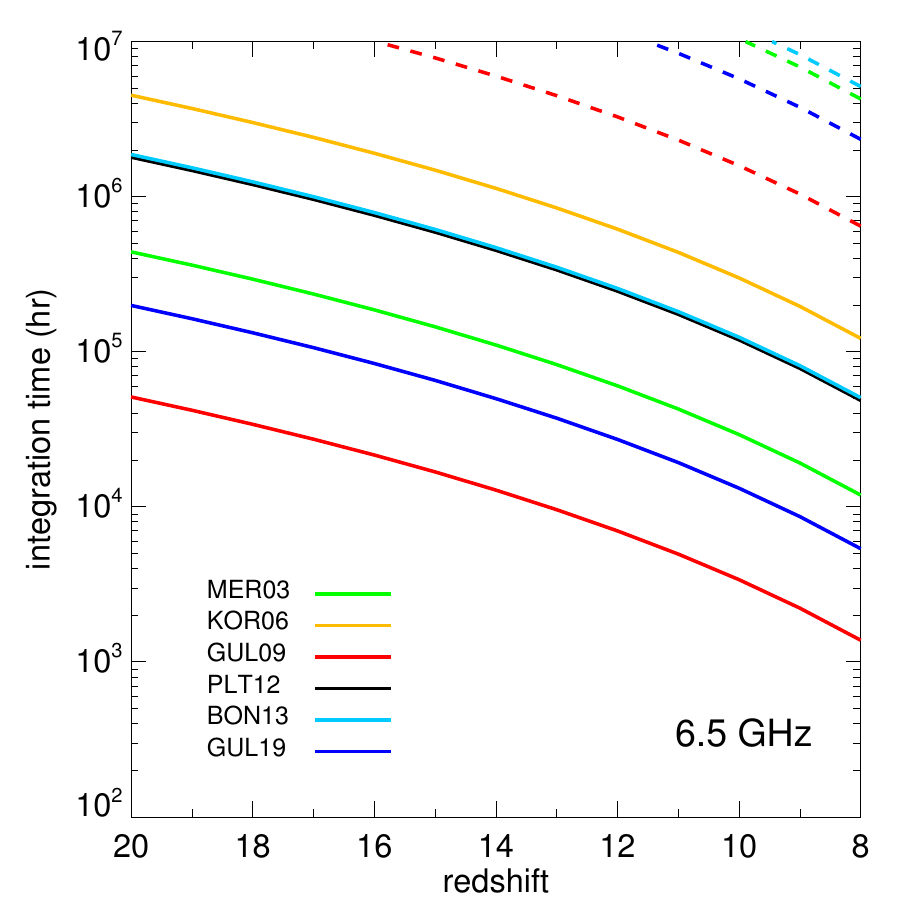,width=0.35\linewidth,clip=}  
\end{tabular}
\end{center}
\caption{Integration times required to detect radio emission from 10$^5$ \Ms\ (dashed) and 10$^6$ \Ms\ (solid) DCBHs at $z =$ 8 - 20.  Upper left:  500 MHz SKA band.  Upper right:  1.5 GHz ngVLA band.  Lower left:  2.5 GHz ngVLA band.  Lower right:  6.5 GHz ngVLA band.}
\vspace{0.1in}
\label{fig:int} 
\end{figure*}

\begin{figure*} 
\begin{center}
\begin{tabular}{cc}
\epsfig{file=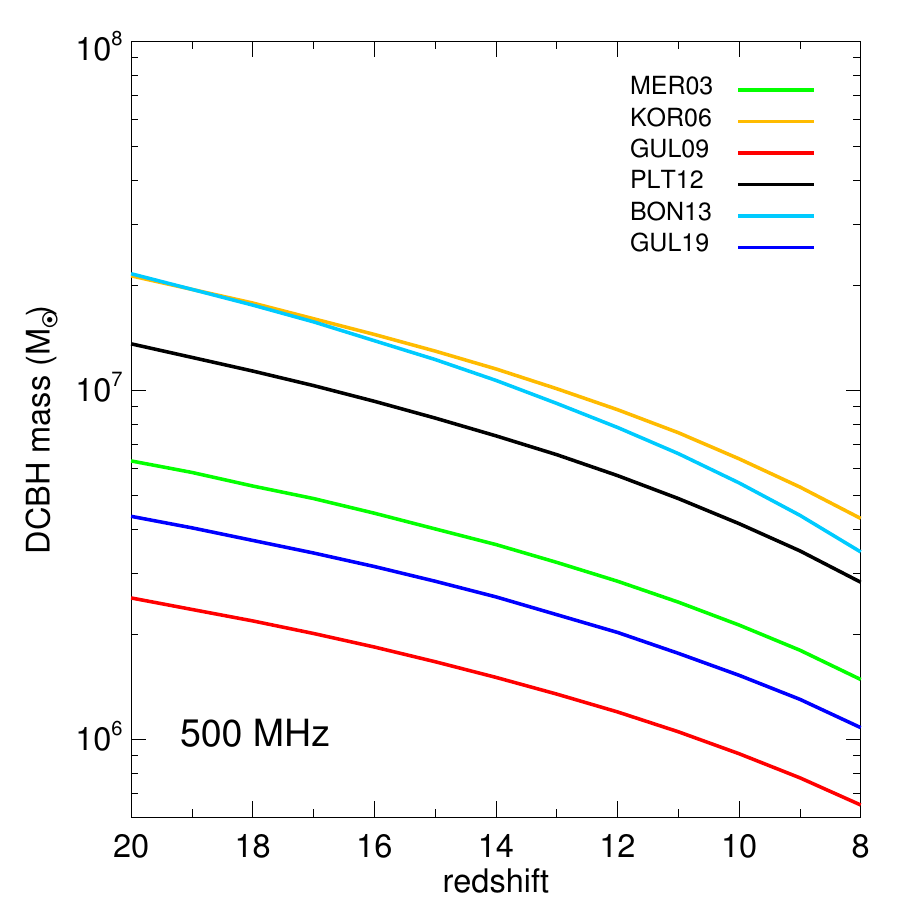,width=0.35\linewidth,clip=}  &
\epsfig{file=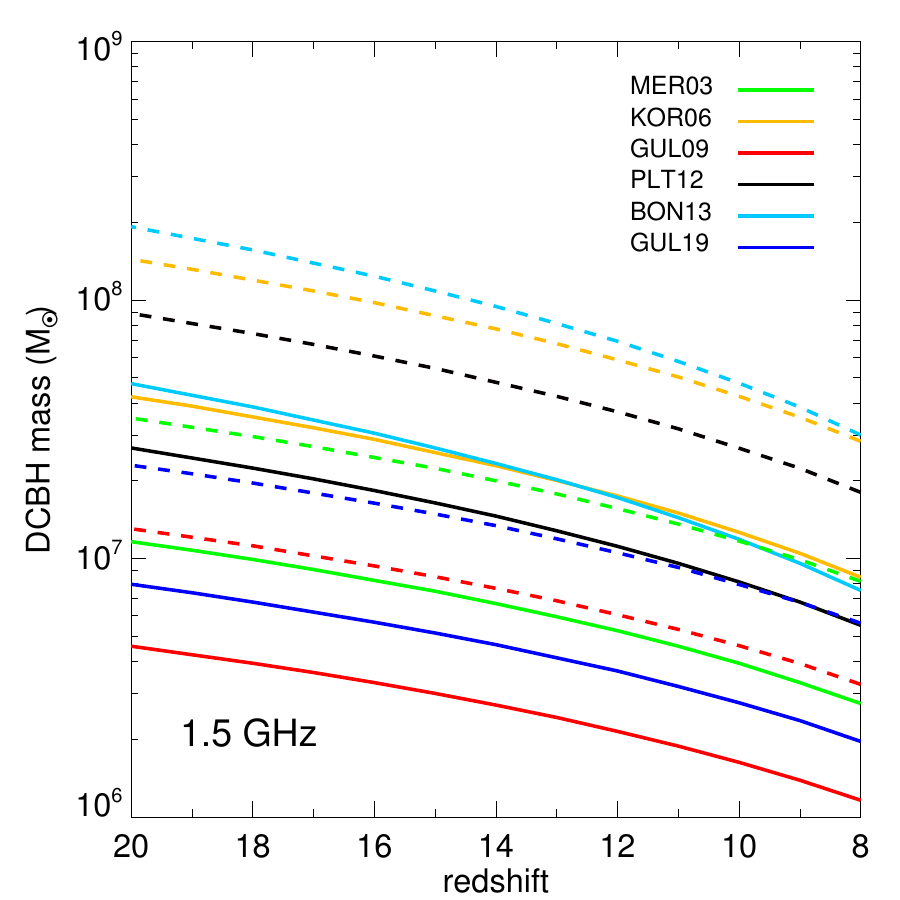,width=0.35\linewidth,clip=}  \\
\epsfig{file=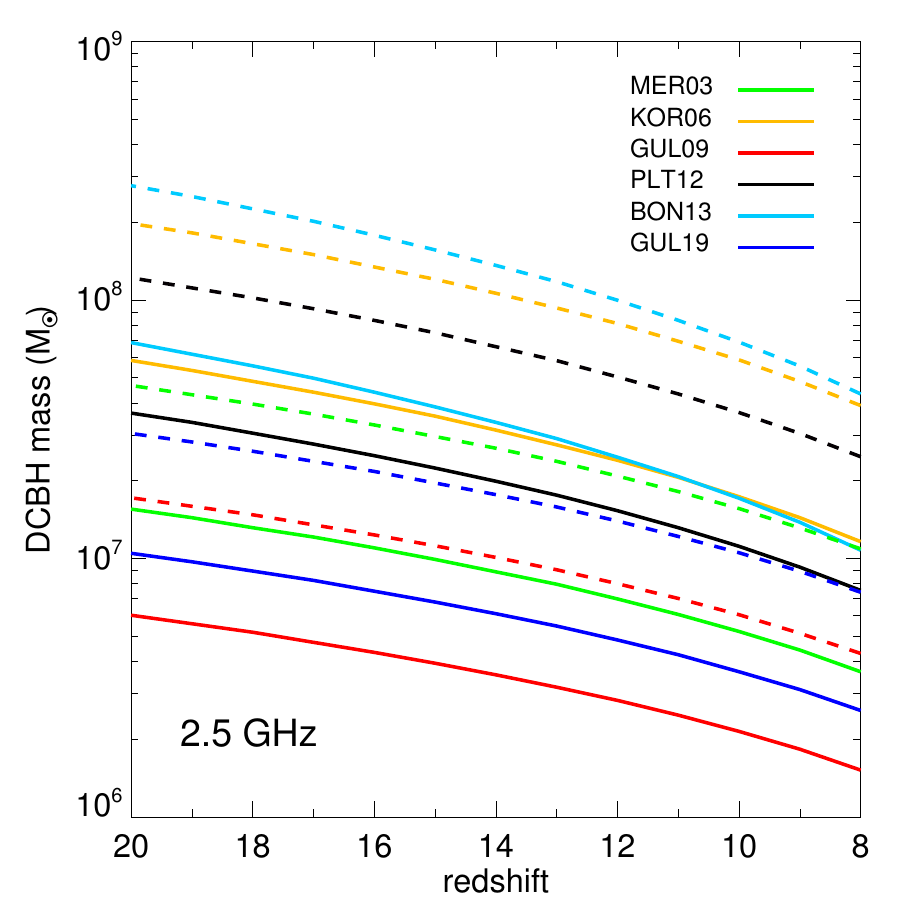,width=0.35\linewidth,clip=}  &
\epsfig{file=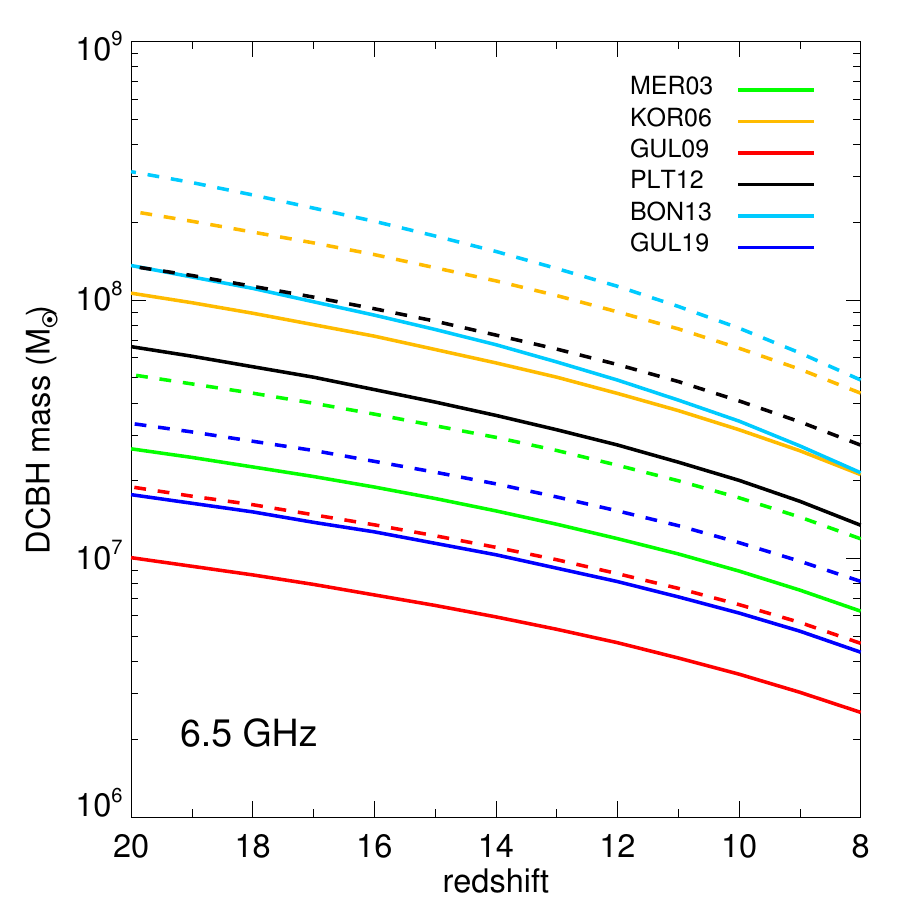,width=0.35\linewidth,clip=}  
\end{tabular}
\end{center}
\caption{Threshold DCBH masses that could be detected in the SKA-FIN survey (solid) and a 24-hr integration time ngVLA survey (dashed) at $z =$ 8 - 20.  Upper left:  500 MHz SKA band.  Upper right:  1.5 GHz ngVLA and SKA bands.  Lower left:  2.5 GHz ngVLA and SKA bands.  Lower right:  6.5 GHz ngVLA and SKA bands.}
\vspace{0.1in}
\label{fig:bhmass} 
\end{figure*}

We note that the supermassive stars that create the 10$^5$ \Ms\ and 10$^6$ \Ms\ BHs considered here probably did not alter flows in their vicinity prior to collapse.  Recent stellar evolution calculations indicate that supermassive primordial stars accreting at average rates above 0.01 \Ms\ yr$^{-1}$ are cool and red over their short lifetimes, producing little ionizing UV radiation that could suppress flows onto the stars \citep{hle18b,herr21a}.  Those accreting at lower rates evolve into hot, blue stars whose ionizing flux could limit flows onto themselves to some degree.  To reach masses of 10$^5$ \Ms\ - 10$^6$ \Ms\ prior to collapse, the stars must accrete at average rates of at least 0.03 \Ms\ yr$^{-1}$ \citep[Figure 4 of][]{tyr17} so they likely evolve as cooler stars.  Our simulations therefore capture the true state of flows onto the DCBH at birth.

\subsection{Fundamental Plane Radio Fluxes}

Fundamental planes of black hole accretion (FPs) are correlations between the mass of a BH, $M_\mathrm{BH}$, its nuclear X-ray luminosity at 2 - 10 keV, $L_\mathrm{X}$, and its nuclear radio luminosity at 5 GHz, $L_\mathrm{R}$ (\citealt{merl03}; see \citealt{mez18} for a brief review).  FPs extend over six orders of magnitude in BH mass, including down to the intermediate mass black hole (IMBH) regime \citep[$< 10^5$ \Ms;][]{gul14}.  To estimate the flux from a DCBH in a given radio band in the observer frame we first use an FP to calculate $L_\mathrm{R}$ in the rest frame, which depends on $M_\mathrm{BH}$ and $L_\mathrm{X}$.  We calculate $L_\mathrm{X}$ from $L_{\mathrm{bol}}$ with Equation 21 of \citet{marc04},
\begin{equation}
\mathrm{log}\left(\frac{L_\mathrm{bol}}{L_\mathrm{X}}\right) = 1.54 + 0.24 \mathcal{L} + 0.012 \mathcal{L}^2 - 0.0015 \mathcal{L}^3,
\end{equation}
where $\mathcal{L} = \mathrm{log} \, L_\mathrm{bol} - 12$ and $L_\mathrm{bol}$ is in units of solar luminosity.  $L_\mathrm{R}$ can then be obtained from $L_\mathrm{X}$ with an FP of the form
\begin{equation}
\mathrm{log} \, L_\mathrm{R} = \alpha \, \mathrm{log} \, L_\mathrm{X} + \beta \, \mathrm{log} \, M_\mathrm{BH} + \gamma,
\end{equation}
where $\alpha$, $\beta$ and $\gamma$ for FPs from \citet[][MER03]{merl03}, \citet[][KOR06]{kord06}, \citet[][GUL09]{gul09}, \citet[][PLT12]{plot12}, and \citet[][BON13]{bonchi13} are listed in Table~1 of \citet{wet20a}.  We also consider the FP of Equation 19 in \citet[][GUL19]{gul19}, 
\begin{equation}
R \, = \, -0.62 + 0.70 \, X + 0.74 \, \mu,
\end{equation}
where $R =$ log($L_\mathrm{R}/10^{38} \mathrm{erg/s}$), $X =$ log($L_\mathrm{X}/10^{40} \mathrm{erg/s}$) and $\mu =$ log($M_\mathrm{BH}/10^{8}$\Ms). 

Since radio flux from a DCBH that is cosmologically redshifted into a given observer band does not originate from 5 GHz in the source frame, we calculate it from $L_\mathrm{R} =$ $\nu L_{\nu}$, assuming that the spectral luminosity $L_{\nu} \propto \nu^{-\alpha}$ with a spectral index $\alpha =$ 0.7 \citep{ccb02}.  The spectral flux at $\nu$ in the observer frame is then determined from the spectral luminosity at $\nu'$ in the rest frame with
\begin{equation}
F_\nu = \frac{L_{\nu'}(1 + z)}{4 \pi {d_\mathrm L}^2},
\end{equation}
where $d_\mathrm L$ is the luminosity distance and $\nu' = (1+z) \nu$.  We plot radio fluxes for 10$^5$ and 10$^6$ \Ms\ DCBHs at 0.85 $L_{\mathrm{Edd}}$ at 500 MHz, 1.5 GHz, 2.5 GHz and 6.5 GHz for all six FPs at $z =$ 8 - 20 in Figure~\ref{fig:flux}.  

\subsection{UC \ion{H}{2} Region Flux}

The UC \ion{H}{2} region of the DCBH in principle can also be a source of synchrotron and bremsstrahlung radio emission that is not included in the FPs, which only consider emission due to the inner accretion disk of the BH and its interaction with its immediate environment.  At a density of $\sim$ 10$^{4}$ cm$^{-3}$, temperature of $\sim$ $5 \times 10^{6}$ K and a maximum radius of 0.5 pc (this radius fluctuates with flows onto the DCBH), the UC \ion{H}{2} region will be optically thick to free-free absorption at rest frequencies below 71 MHz. Consequently, it will shine like a black body at $T =$ $5 \times 10^{6}$ K at these frequencies.  For emission at frequency $\nu'$ by an \ion{H}{2} region of radius $R$ at redshift $z$, the received flux at frequency $\nu  = \nu'/ (1+z)$ is 
\begin{equation}
F_{\nu} = \frac{2 \pi kT}{{\lambda'}^2} (1 + z) \frac{R^2}{{d_\mathrm L}^2},
\end{equation}
where $\lambda' = c / \nu'$.  For $z =$ 9 and our cosmological parameters, this becomes 
\begin{equation}
F_{\nu} = 1.4 \times 10^{-5} \, \mathrm{nJy} \left(\frac{\nu'}{1 \, \mathrm{MHz}} \right)^2.
\end{equation}
Besides lying well below SKA and ngVLA bands, these fluxes would be far too low to be observed, and they represent an upper limit because the UC \ion{H}{2} region would be optically thin to radio frequencies above 71 MHz and would thus likely be  dimmer at the frequencies considered in this paper. 

\section{DCBH Radio Power}

The SKA-FIN all-sky survey will reach 20 nJy at all 4 frequencies considered here and the ngVLA can reach 45 nJy at 3.5 - 12.3 GHz and 78 nJy at 1.2 - 3.5 GHz with 24 hr integration times \citep{pr18}.  Our calculations show that 10$^6$ \Ms\ DCBHs could be found by the SKA-FIN survey at 500 MHz out to $z =$ 11 but detections by the ngVLA at 1.5 - 6.5 MHz will require integration times greater than 24 hr.

We plot these integration times for $z =$ 8 - 20 in Figure~\ref{fig:int}.  With total exposures of 1000 - 2000 hr, ngVLA could detect 10$^6$ \Ms\ DCBHs at $z =$ 8 - 10.  This is the redshift range over which some models predict that the comoving number density of DCBHs is highest  \citep[see, e.g., Figure~4 of][]{rosa17} so both observatories can detect these objects at the times at which their numbers are greatest.  However, we find that DCBHs with masses $\lesssim$ 10$^5$ \Ms\ will remain beyond the reach of these telescopes for now.  

In Figure~\ref{fig:bhmass} we show the minimum DCBH mass required for detection by the SKA-FIN survey and by the ngVLA with a 24-hr integration time.  The least massive one that could be detected by currently proposed surveys would be a 6.5 $\times$ 10$^5$ \Ms\ DCBH at $z =$ 8 at 500 MHz by SKA-FIN.  However, this survey could find BHs ranging in mass from 2.5 $\times$ 10$^6$ \Ms\ to 2.2 $\times$ 10$^7$ \Ms\ at $z =$ 20 and ngVLA could detect 10$^7$ - 10$^8$ \Ms\ BHs at $z =$ 15 - 20.  Such detections at high redshifts could probe the origins of the first quasars and distinguish between channels for their formation, such as Pop III stars or very massive stars formed in intermediate Lyman-Werner UV backgrounds \citep{pat21b}.

\section{Discussion and Conclusion}

The SKA could detect DCBHs at birth at $z =$ 8 - 9, when their comoving number densities may be highest, but only more massive, evolved BHs could be detected at higher redshifts \citep{wet21b}.  Radio flux from these BHs may be contaminated by emission from H II regions and SNRs in their host galaxies, which would host active star formation \citep{smidt18}.  However, \citet{wet20a} found that in CR7, the brightest Ly$\alpha$ emitter at $z >$ 6 and which hosts rapid star formation, these competing sources of radio flux were either too small to be detected or could be distinguished from flux from a 4 $\times$ 10$^6$ \Ms\ DCBH.  \citet{mez19} detected IMBHs in dwarf galaxies by subtracting contributions to their radio signals from these additional sources.  Here, we only consider the most massive DCBHs forming in purely atomically-cooling halos \citep[e.g.,][]{pat21a}, but less massive objects forming in lower LW backgrounds (10$^3$ \Ms\ - 10$^4$ \Ms) may have been more common in the primordial universe \citep{latif20d}.  Our calculations show that these IMBHs would not be visible at birth and would have to grow in mass by at least a factor of ten to be found at later epochs.

Our radio fluxes ignore contributions from DCBH jets but they are not thought to be important for two reasons.  First, as discussed earlier, jets are not expected at the accretion rates we found for DCBHs at birth.  Second, even if a DCBH launches a jet, its radio emission may be quenched by the CMB at high redshifts.  If the energy density of CMB photons (which rises as $z^4$) exceeds the magnetic field energy density in the lobes of the jet, relativistic electrons preferentially cool by upscattering CMB photons instead of synchrotron radiation.  The absence of radio emission from some high-redshift quasars has been attributed to this process \citep[e.g.,][]{fg14}.  

The SKA and ngVLA could, in synergy with {\em JWST}, {\em Euclid}, and the {\em RST}, probe the origins and numbers of the first quasars because they could detect BHs that are a few 10$^6$ - 10$^7$ \Ms\ out to $z =$ 20 in the coming decade.  Redshift cutoffs above which such objects are not observed in the radio could distinguish  between seeding mechanisms for the first SMBHs, whether they come from ordinary Pop III stars, supermassive primordial stars, or are the result of runaway stellar collisions in dense nuclear clusters at high redshifts.  However, modeling the radio emission from this primordial population of low-mass quasars would require cosmological simulations of star formation rates in their host galaxies to discriminate between competing sources of radio emission in them, such as SNRs and \ion{H}{2} regions.  These models are now under development.

\acknowledgments

D.J.W. was supported by the Ida Pfeiffer Professorship at the Institute of Astrophysics at the University of Vienna.  MM acknowledges support from the Beatriu de Pinos fellowship (2017-BP-00114) and from the Ramon y Cajal fellowship (RYC2019-027670-I).  S.P. was supported by STFC grant ST/N504245/1.  A. M. acknowledges support from the UK Science and Technology Facilities Council Consolidated Grant ST/R000972/1. M.L. acknowledges funding from UAEU via UPAR grant No. 31S372.  Our calculations were performed on the Sciama HPC cluster at the Institute of Cosmology and Gravitation at the University of Portsmouth.


\begin{thebibliography}{}
\expandafter\ifx\csname natexlab\endcsname\relax\def\natexlab#1{#1}\fi
\providecommand{\url}[1]{\href{#1}{#1}}
\providecommand{\dodoi}[1]{doi:~\href{http://doi.org/#1}{\nolinkurl{#1}}}
\providecommand{\doeprint}[1]{\href{http://ascl.net/#1}{\nolinkurl{http://ascl.net/#1}}}
\providecommand{\doarXiv}[1]{\href{https://arxiv.org/abs/#1}{\nolinkurl{https://arxiv.org/abs/#1}}}

\bibitem[{{Ba{\~n}ados} {et~al.}(2018){Ba{\~n}ados}, {Venemans},
  {Mazzucchelli}, {Farina}, {Walter}, {Wang}, {Decarli}, {Stern}, {Fan},
  {Davies}, {Hennawi}, {Simcoe}, {Turner}, {Rix}, {Yang}, {Kelson}, {Rudie}, \&
  {Winters}}]{ban18}
{Ba{\~n}ados}, E., {Venemans}, B.~P., {Mazzucchelli}, C., {et~al.} 2018, \nat,
  553, 473, \dodoi{10.1038/nature25180}

\bibitem[{{Barrow} {et~al.}(2018){Barrow}, {Aykutalp}, \& {Wise}}]{bar18}
{Barrow}, K. S.~S., {Aykutalp}, A., \& {Wise}, J.~H. 2018, Nature Astronomy, 2,
  987, \dodoi{10.1038/s41550-018-0569-y}

\bibitem[{{Bonchi} {et~al.}(2013){Bonchi}, {La Franca}, {Melini}, {Bongiorno},
  \& {Fiore}}]{bonchi13}
{Bonchi}, A., {La Franca}, F., {Melini}, G., {Bongiorno}, A., \& {Fiore}, F.
  2013, \mnras, 429, 1970, \dodoi{10.1093/mnras/sts456}

\bibitem[{{Bryan} {et~al.}(2014){Bryan}, {Norman}, {O'Shea}, {Abel}, {Wise},
  {Turk}, {Reynolds}, {Collins}, {Wang}, {Skillman}, {Smith}, {Harkness},
  {Bordner}, {Kim}, {Kuhlen}, {Xu}, {Goldbaum}, {Hummels}, {Kritsuk}, {Tasker},
  {Skory}, {Simpson}, {Hahn}, {Oishi}, {So}, {Zhao}, {Cen}, {Li}, \& {Enzo
  Collaboration}}]{enzo}
{Bryan}, G.~L., {Norman}, M.~L., {O'Shea}, B.~W., {et~al.} 2014, \apjs, 211,
  19, \dodoi{10.1088/0067-0049/211/2/19}

\bibitem[{{Condon}(1992)}]{con92}
{Condon}, J.~J. 1992, \araa, 30, 575,
  \dodoi{10.1146/annurev.aa.30.090192.003043}

\bibitem[{{Condon} {et~al.}(2002){Condon}, {Cotton}, \& {Broderick}}]{ccb02}
{Condon}, J.~J., {Cotton}, W.~D., \& {Broderick}, J.~J. 2002, \aj, 124, 675,
  \dodoi{10.1086/341650}

\bibitem[{{DeBuhr} {et~al.}(2010){DeBuhr}, {Quataert}, {Ma}, \&
  {Hopkins}}]{alphad}
{DeBuhr}, J., {Quataert}, E., {Ma}, C.-P., \& {Hopkins}, P. 2010, \mnras, 406,
  L55, \dodoi{10.1111/j.1745-3933.2010.00881.x}

\bibitem[{{Fabian} {et~al.}(2014){Fabian}, {Walker}, {Celotti}, {Ghisellini},
  {Mocz}, {Blundell}, \& {McMahon}}]{fg14}
{Fabian}, A.~C., {Walker}, S.~A., {Celotti}, A., {et~al.} 2014, \mnras, 442,
  L81, \dodoi{10.1093/mnrasl/slu065}

\bibitem[{{G{\"u}ltekin} {et~al.}(2014){G{\"u}ltekin}, {Cackett}, {King},
  {Miller}, \& {Pinkney}}]{gul14}
{G{\"u}ltekin}, K., {Cackett}, E.~M., {King}, A.~L., {Miller}, J.~M., \&
  {Pinkney}, J. 2014, \apjl, 788, L22, \dodoi{10.1088/2041-8205/788/2/L22}

\bibitem[{{G{\"u}ltekin} {et~al.}(2009){G{\"u}ltekin}, {Cackett}, {Miller}, {Di
  Matteo}, {Markoff}, \& {Richstone}}]{gul09}
{G{\"u}ltekin}, K., {Cackett}, E.~M., {Miller}, J.~M., {et~al.} 2009, \apj,
  706, 404, \dodoi{10.1088/0004-637X/706/1/404}

\bibitem[{{G{\"u}ltekin} {et~al.}(2019){G{\"u}ltekin}, {King}, {Cackett},
  {Nyland}, {Miller}, {Di Matteo}, {Markoff}, \& {Rupen}}]{gul19}
{G{\"u}ltekin}, K., {King}, A.~L., {Cackett}, E.~M., {et~al.} 2019, \apj, 871,
  80, \dodoi{10.3847/1538-4357/aaf6b9}

\bibitem[{{Haemmerl{\'e}} {et~al.}(2018){Haemmerl{\'e}}, {Woods}, {Klessen},
  {Heger}, \& {Whalen}}]{hle18b}
{Haemmerl{\'e}}, L., {Woods}, T.~E., {Klessen}, R.~S., {Heger}, A., \&
  {Whalen}, D.~J. 2018, \mnras, 474, 2757, \dodoi{10.1093/mnras/stx2919}

\bibitem[{{Hahn} \& {Abel}(2011)}]{hahn11}
{Hahn}, O., \& {Abel}, T. 2011, \mnras, 415, 2101,
  \dodoi{10.1111/j.1365-2966.2011.18820.x}

\bibitem[{{Herrington} {et~al.}(2021){Herrington}, {Whalen}, \&
  {Wood}}]{herr21a}
{Herrington}, N., {Whalen}, D.~J., \& {Wood}, T.~E. 2021

\bibitem[{{Johnson} {et~al.}(2013){Johnson}, {Whalen}, {Li}, \& {Holz}}]{jet13}
{Johnson}, J.~L., {Whalen}, D.~J., {Li}, H., \& {Holz}, D.~E. 2013, \apj, 771,
  116, \dodoi{10.1088/0004-637X/771/2/116}

\bibitem[{{K{\"o}rding} {et~al.}(2006){K{\"o}rding}, {Falcke}, \&
  {Corbel}}]{kord06}
{K{\"o}rding}, E., {Falcke}, H., \& {Corbel}, S. 2006, \aap, 456, 439,
  \dodoi{10.1051/0004-6361:20054144}

\bibitem[{{Latif} \& {Khochfar}(2020)}]{latif20b}
{Latif}, M.~A., \& {Khochfar}, S. 2020, \mnras, 497, 3761,
  \dodoi{10.1093/mnras/staa2218}

\bibitem[{{Latif} {et~al.}(2020){Latif}, {Khochfar}, {Schleicher}, \&
  {Whalen}}]{latif20d}
{Latif}, M.~A., {Khochfar}, S., {Schleicher}, D., \& {Whalen}, D.~J. 2020,
  arXiv e-prints, arXiv:2012.09177.
\newblock \doarXiv{2012.09177}

\bibitem[{{Marconi} {et~al.}(2004){Marconi}, {Risaliti}, {Gilli}, {Hunt},
  {Maiolino}, \& {Salvati}}]{marc04}
{Marconi}, A., {Risaliti}, G., {Gilli}, R., {et~al.} 2004, \mnras, 351, 169,
  \dodoi{10.1111/j.1365-2966.2004.07765.x}

\bibitem[{{Matsuoka} {et~al.}(2019){Matsuoka}, {Onoue}, {Kashikawa}, {Strauss},
  {Iwasawa}, {Lee}, {Imanishi}, {Nagao}, {Akiyama}, {Asami}, {Bosch},
  {Furusawa}, {Goto}, {Gunn}, {Harikane}, {Ikeda}, {Izumi}, {Kawaguchi},
  {Kato}, {Kikuta}, {Kohno}, {Komiyama}, {Koyama}, {Lupton}, {Minezaki},
  {Miyazaki}, {Murayama}, {Niida}, {Nishizawa}, {Noboriguchi}, {Oguri}, {Ono},
  {Ouchi}, {Price}, {Sameshima}, {Schulze}, {Shirakata}, {Silverman},
  {Sugiyama}, {Tait}, {Takada}, {Takata}, {Tanaka}, {Tang}, {Toba}, {Utsumi},
  {Wang}, \& {Yamashita}}]{mats19}
{Matsuoka}, Y., {Onoue}, M., {Kashikawa}, N., {et~al.} 2019, \apjl, 872, L2,
  \dodoi{10.3847/2041-8213/ab0216}

\bibitem[{{Meiksin} \& {Whalen}(2013)}]{mw12}
{Meiksin}, A., \& {Whalen}, D.~J. 2013, \mnras, 430, 2854,
  \dodoi{10.1093/mnras/stt089}

\bibitem[{{Merloni} \& {Heinz}(2008)}]{mh08}
{Merloni}, A., \& {Heinz}, S. 2008, \mnras, 388, 1011,
  \dodoi{10.1111/j.1365-2966.2008.13472.x}

\bibitem[{{Merloni} {et~al.}(2003){Merloni}, {Heinz}, \& {di Matteo}}]{merl03}
{Merloni}, A., {Heinz}, S., \& {di Matteo}, T. 2003, \mnras, 345, 1057,
  \dodoi{10.1046/j.1365-2966.2003.07017.x}

\bibitem[{{Mezcua} {et~al.}(2018){Mezcua}, {Hlavacek-Larrondo}, {Lucey},
  {Hogan}, {Edge}, \& {McNamara}}]{mez18}
{Mezcua}, M., {Hlavacek-Larrondo}, J., {Lucey}, J.~R., {et~al.} 2018, \mnras,
  474, 1342, \dodoi{10.1093/mnras/stx2812}

\bibitem[{{Mezcua} {et~al.}(2019){Mezcua}, {Suh}, \& {Civano}}]{mez19}
{Mezcua}, M., {Suh}, H., \& {Civano}, F. 2019, \mnras, 488, 685,
  \dodoi{10.1093/mnras/stz1760}

\bibitem[{{Mortlock} {et~al.}(2011){Mortlock}, {Warren}, {Venemans}, {Patel},
  {Hewett}, {McMahon}, {Simpson}, {Theuns}, {Gonz{\'a}les-Solares}, {Adamson},
  {Dye}, {Hambly}, {Hirst}, {Irwin}, {Kuiper}, {Lawrence}, \&
  {R{\"o}ttgering}}]{mort11}
{Mortlock}, D.~J., {Warren}, S.~J., {Venemans}, B.~P., {et~al.} 2011, \nat,
  474, 616, \dodoi{10.1038/nature10159}

\bibitem[{{Pacucci} {et~al.}(2015){Pacucci}, {Ferrara}, {Volonteri}, \&
  {Dubus}}]{pac15}
{Pacucci}, F., {Ferrara}, A., {Volonteri}, M., \& {Dubus}, G. 2015, \mnras,
  454, 3771, \dodoi{10.1093/mnras/stv2196}

\bibitem[{{Patrick} {et~al.}(2021){Patrick}, {Whalen}, {Elford}, \&
  {Latif}}]{pat21b}
{Patrick}, S., {Whalen}, D.~J., {Elford}, J.~S., \& {Latif}, M. 2021

\bibitem[{{Patrick} {et~al.}(2020){Patrick}, {Whalen}, {Elford}, \&
  {Latif}}]{pat21a}
{Patrick}, S.~J., {Whalen}, D.~J., {Elford}, J.~S., \& {Latif}, M.~A. 2020,
  arXiv e-prints, arXiv:2012.11612.
\newblock \doarXiv{2012.11612}

\bibitem[{{Planck Collaboration} {et~al.}(2016){Planck Collaboration}, {Ade},
  {Aghanim}, {Arnaud}, {Ashdown}, {Aumont}, {Baccigalupi}, {Banday},
  {Barreiro}, {Bartlett}, \& et~al.}]{planck2}
{Planck Collaboration}, {Ade}, P.~A.~R., {Aghanim}, N., {et~al.} 2016, \aap,
  594, A13, \dodoi{10.1051/0004-6361/201525830}

\bibitem[{{Plotkin} {et~al.}(2012){Plotkin}, {Markoff}, {Kelly}, {K{\"o}rding},
  \& {Anderson}}]{plot12}
{Plotkin}, R.~M., {Markoff}, S., {Kelly}, B.~C., {K{\"o}rding}, E., \&
  {Anderson}, S.~F. 2012, \mnras, 419, 267,
  \dodoi{10.1111/j.1365-2966.2011.19689.x}

\bibitem[{{Plotkin} \& {Reines}(2018)}]{pr18}
{Plotkin}, R.~M., \& {Reines}, A.~E. 2018, arXiv:1810.06814, arXiv:1810.06814.
\newblock \doarXiv{1810.06814}

\bibitem[{{Smidt} {et~al.}(2018){Smidt}, {Whalen}, {Johnson}, {Surace}, \&
  {Li}}]{smidt18}
{Smidt}, J., {Whalen}, D.~J., {Johnson}, J.~L., {Surace}, M., \& {Li}, H. 2018,
  \apj, 865, 126, \dodoi{10.3847/1538-4357/aad7b8}

\bibitem[{{Smith} {et~al.}(2018){Smith}, {Regan}, {Downes}, {Norman}, {O'Shea},
  \& {Wise}}]{srd18}
{Smith}, B.~D., {Regan}, J.~A., {Downes}, T.~P., {et~al.} 2018, \mnras, 480,
  3762, \dodoi{10.1093/mnras/sty2103}

\bibitem[{{Surace} {et~al.}(2019){Surace}, {Zackrisson}, {Whalen}, {Hartwig},
  {Glover}, {Woods}, {Heger}, \& {Glover}}]{sur19a}
{Surace}, M., {Zackrisson}, E., {Whalen}, D.~J., {et~al.} 2019, \mnras, 488,
  3995, \dodoi{10.1093/mnras/stz1956}

\bibitem[{{Surace} {et~al.}(2018){Surace}, {Whalen}, {Hartwig}, {Zackrisson},
  {Glover}, {Patrick}, {Woods}, {Heger}, \& {Haemmerl{\'e}}}]{sur18a}
{Surace}, M., {Whalen}, D.~J., {Hartwig}, T., {et~al.} 2018, \apjl, 869, L39,
  \dodoi{10.3847/2041-8213/aaf80d}

\bibitem[{{Umeda} {et~al.}(2016){Umeda}, {Hosokawa}, {Omukai}, \&
  {Yoshida}}]{um16}
{Umeda}, H., {Hosokawa}, T., {Omukai}, K., \& {Yoshida}, N. 2016, \apjl, 830,
  L34, \dodoi{10.3847/2041-8205/830/2/L34}

\bibitem[{{Valiante} {et~al.}(2017){Valiante}, {Agarwal}, {Habouzit}, \&
  {Pezzulli}}]{rosa17}
{Valiante}, R., {Agarwal}, B., {Habouzit}, M., \& {Pezzulli}, E. 2017, \pasa,
  34, e031, \dodoi{10.1017/pasa.2017.25}

\bibitem[{{Vikaeus} {et~al.}(2021){Vikaeus}, {Zackrisson}, \&
  {Whalen}}]{vik21a}
{Vikaeus}, A.~F., {Zackrisson}, E., \& {Whalen}, D.~J. 2021

\bibitem[{{Wang} {et~al.}(2021){Wang}, {Yang}, {Fan}, {Hennawi}, {Barth},
  {Banados}, {Bian}, {Boutsia}, {Connor}, {Davies}, {Decarli}, {Eilers},
  {Farina}, {Green}, {Jiang}, {Li}, {Mazzucchelli}, {Nanni}, {Schindler},
  {Venemans}, {Walter}, {Wu}, \& {Yue}}]{wang21}
{Wang}, F., {Yang}, J., {Fan}, X., {et~al.} 2021, \apjl, 907, L1,
  \dodoi{10.3847/2041-8213/abd8c6}

\bibitem[{{Whalen} {et~al.}(2004){Whalen}, {Abel}, \& {Norman}}]{wan04}
{Whalen}, D., {Abel}, T., \& {Norman}, M.~L. 2004, \apj, 610, 14,
  \dodoi{10.1086/421548}

\bibitem[{{Whalen} \& {Fryer}(2012)}]{wf12}
{Whalen}, D.~J., \& {Fryer}, C.~L. 2012, \apjl, 756, L19,
  \dodoi{10.1088/2041-8205/756/1/L19}

\bibitem[{{Whalen} \& {Mezcua}(2021)}]{wet21b}
{Whalen}, D.~J., \& {Mezcua}, M. 2021

\bibitem[{{Whalen} {et~al.}(2020{\natexlab{a}}){Whalen}, {Mezcua}, {Meiksin},
  {Hartwig}, \& {Latif}}]{wet20a}
{Whalen}, D.~J., {Mezcua}, M., {Meiksin}, A., {Hartwig}, T., \& {Latif}, M.~A.
  2020{\natexlab{a}}, \apjl, 896, L45, \dodoi{10.3847/2041-8213/ab9a30}

\bibitem[{{Whalen} {et~al.}(2020{\natexlab{b}}){Whalen}, {Surace}, {Bernhardt},
  {Zackrisson}, {Pacucci}, {Ziegler}, \& {Hirschmann}}]{wet20b}
{Whalen}, D.~J., {Surace}, M., {Bernhardt}, C., {et~al.} 2020{\natexlab{b}},
  \apjl, 897, L16, \dodoi{10.3847/2041-8213/ab9d29}

\bibitem[{{Wise} \& {Abel}(2011)}]{moray}
{Wise}, J.~H., \& {Abel}, T. 2011, \mnras, 414, 3458,
  \dodoi{10.1111/j.1365-2966.2011.18646.x}

\bibitem[{{Woods} {et~al.}(2017){Woods}, {Heger}, {Whalen}, {Haemmerl{\'e}}, \&
  {Klessen}}]{tyr17}
{Woods}, T.~E., {Heger}, A., {Whalen}, D.~J., {Haemmerl{\'e}}, L., \&
  {Klessen}, R.~S. 2017, \apjl, 842, L6, \dodoi{10.3847/2041-8213/aa7412}

\bibitem[{{Woods} {et~al.}(2021){Woods}, {Patrick}, {Elford}, {Whalen}, \&
  {Heger}}]{tyr21a}
{Woods}, T.~E., {Patrick}, S., {Elford}, J.~S., {Whalen}, D.~J., \& {Heger}, A.
  2021, \apj, 915, 110, \dodoi{10.3847/1538-4357/abfaf9}

\bibitem[{{Woods} {et~al.}(2019){Woods}, {Agarwal}, {Bromm}, {Bunker}, {Chen},
  {Chon}, {Ferrara}, {Glover}, {Haemmerl{\'e}}, {Haiman}, {Hartwig}, {Heger},
  {Hirano}, {Hosokawa}, {Inayoshi}, {Klessen}, {Kobayashi}, {Koliopanos},
  {Latif}, {Li}, {Mayer}, {Mezcua}, {Natarajan}, {Pacucci}, {Rees}, {Regan},
  {Sakurai}, {Salvadori}, {Schneider}, {Surace}, {Tanaka}, {Whalen}, \&
  {Yoshida}}]{titans}
{Woods}, T.~E., {Agarwal}, B., {Bromm}, V., {et~al.} 2019, Publications of the
  Astronomical Society of Australia, 36, e027, \dodoi{10.1017/pasa.2019.14}

\bibitem[{{Yue} \& {Ferrara}(2021)}]{yf21}
{Yue}, B., \& {Ferrara}, A. 2021, arXiv e-prints, arXiv:2107.11307.
\newblock \doarXiv{2107.11307}

\end{thebibliography}

\end{document}